\documentclass[%
 aip,
 amsmath,amssymb,
 reprint,%
]{revtex4-1}

\usepackage{graphicx}% Include figure files
\usepackage{dcolumn}% Align table columns on decimal point
\usepackage{bm}% bold math
\usepackage[utf8]{inputenc}
\usepackage[T1]{fontenc}
\usepackage{mathptmx}
\usepackage{etoolbox}

\newcommand{\abs}[1]{\left| #1 \right|}

\preprint{AIP/123-QED}

\makeatletter
\def\@email#1#2{%
 \endgroup
 \patchcmd{\titleblock@produce}
  {\frontmatter@RRAPformat}
  {\frontmatter@RRAPformat{\produce@RRAP{*#1\href{mailto:#2}{#2}}}\frontmatter@RRAPformat}
  {}{}
}%
\makeatother

\begin{document}

\title{Few-MHz bandwidth tunable optical filter based on a fiber-ring resonator} %Title of paper

\author{Gabriele Maron}
\author{Anton Bölian}
\author{Xin-Xin Hu}
\author{Luke Masters}
\author{Arno Rauschenbeutel}
    \email{arno.rauschenbeutel@hu-berlin.de}
\author{J\"urgen Volz}
    \email{juergen.volz@hu-berlin.de}
\affiliation{Department of Physics, Humboldt Universit\"at zu Berlin, 10099 Berlin, Germany}

%\date{\today}

\begin{abstract}
We present a fiber-ring resonator that realizes an ultra-narrowband, high-extinction, low-loss, tunable optical filter. It consists of a pair of commercial variable ratio directional couplers that allow precise adjustment of the filter bandwidth and its on-resonance transmission. This design also grants access to multiple modes of operation, such as a simultaneous band-stop and band-pass filter. Our characterization reveals a bandwidth of less than $2$ MHz, together with an extinction exceeding $20$~dB.
The tunability of the filter properties establishes our device as a versatile platform for selective frequency filtering with sub-natural atomic linewidth resolution.
\end{abstract}


\maketitle 

The use of optical filters is common in most optical applications, allowing the spectral selection or rejection of a particular frequency band and the splitting or combining of light fields of different wavelengths.
A plethora of general as well as application-specific commercial filters are readily available and are often engineered to offer high transmittance or reflectance over narrow wavelength ranges. \cite{Bass1995,Macleod} 
However, the achievable bandwidth of commercial filters based on thin-film interference is typically restricted to a few nanometers by inhomogeneities in the thickness of the films. \cite{giacomo59} 
Moreover, a smaller bandwidth generally comes at the cost of a reduced peak transmission/reflection. \cite{Koonen2006}
Among the available alternatives, filters based on atomic vapors reduce the bandwidth to the GHz range but are fundamentally limited by the specific properties of the atomic medium and by collisional broadening effects. \cite{Xue2012, Widmann2018}
Filters based on Fabry-P\'erot or ring resonators further improve the key filtering parameters compared to these designs, with commercial devices typically reaching a bandwidth of several MHz. \cite{palittapongarnpim12, ahlrichs13, phillips20, cordier23}
For this reason, filter resonators have become key components in optical communication applications, aiding in the generation, processing, and measurement of optical signals, from laser mode selection and stabilization \cite{Reed,Fox2003} to wavelength division multiplexing \cite{Yoo1996, Hou2023} and spectroscopy. \cite{Sun2022, Tong2024}
Despite their proven efficacy, the performance of modular resonators is highly sensitive to the precise alignment of their constituent components. Additionally, free-space light coupling requires careful optimization to ensure accurate mode matching with the cavity.
In contrast, fiber-based resonators alleviate these drawbacks by providing high spectral resolution and transmission (or reflection) efficiency while maintaining simple control over the filter characteristics. 

\begin{figure}[hb]
\centering
    \includegraphics[width=0.48\textwidth,keepaspectratio]{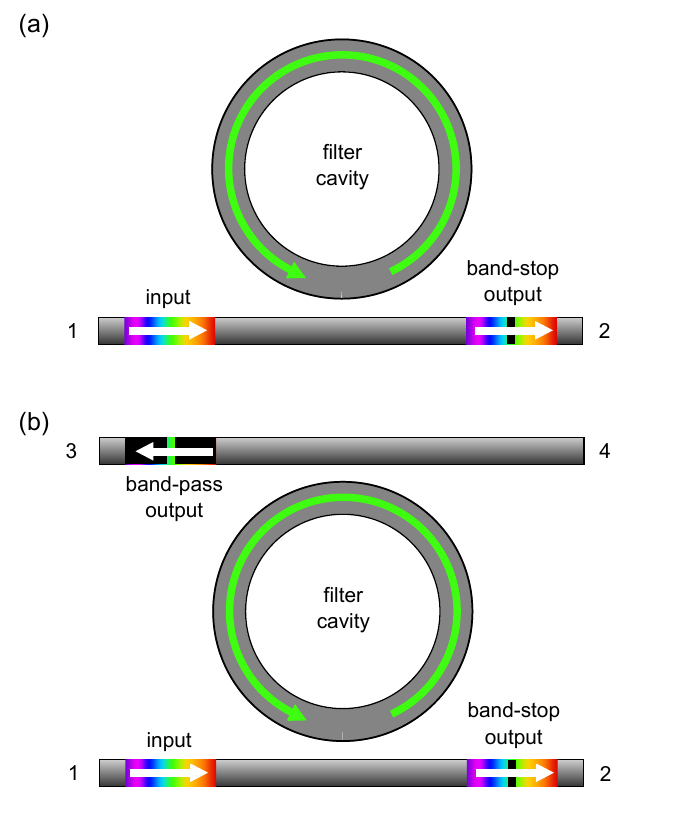}
    \caption{\label{figCartoonSketch} (a) Filter design operating in a two-port configuration as a band-stop filter. Port 1 is used to inject light, and the band-stop filtered light exits through port 2.
    (b) Alternatively, the filter can be operated in a four-port add-drop configuration as a simultaneous band-stop and band-pass filter.
    In this case, the band-pass filtered light exits through port 3, while the band-stop filtered light continues to exit through port 2.}
\end{figure}

\begin{figure*}[t]
    \centering
    \includegraphics[width=0.92\textwidth,keepaspectratio]{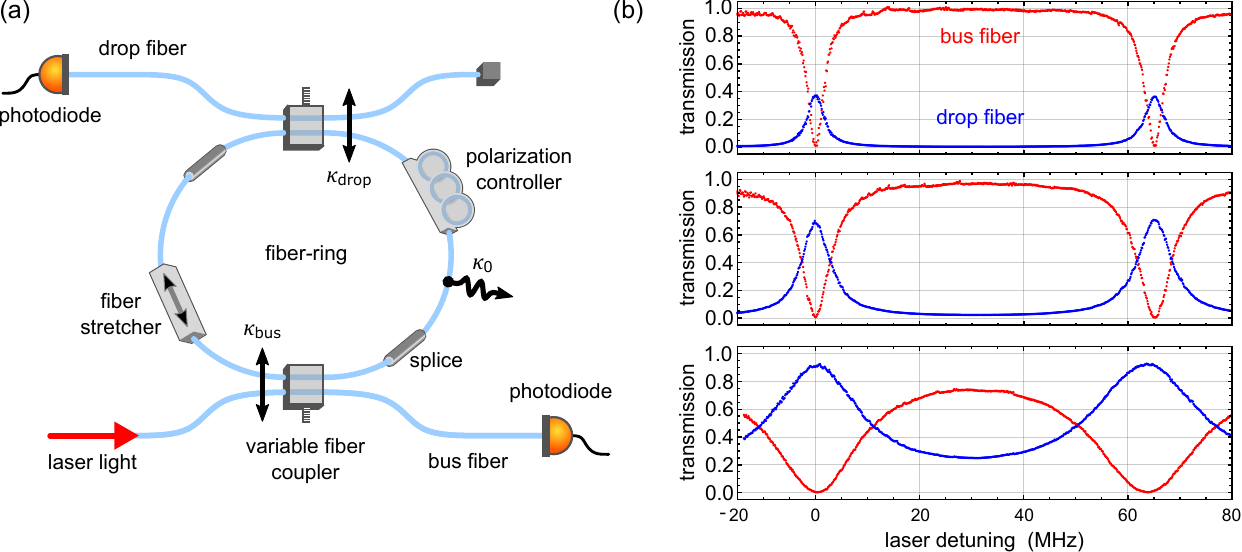}
    \caption{\label{figExpSetup} 
    (a) Add-drop FRR filter setup. Laser light ($\lambda\sim852$ nm), injected into the bus fiber, is coupled into the fiber-ring at a rate $\kappa_\textrm{bus}$, set by a variable ratio directional coupler. A second variable ratio directional coupler couples light from the ring resonator to the drop port at a rate $\kappa_\textrm{drop}$. The transmission in each fiber output is measured using a photodiode. The FRR includes a fiber stretcher that allows us to tune the resonance frequency, as well as a polarization controller to adjust the frequency splitting between the two orthogonally polarized resonator eigenmodes.
    The intrinsic loss rate $\kappa_\textrm{0}$ of the FRR is dominated by the losses at the two fiber splices in the ring and by the internal losses of the variable ratio directional couplers.
    (b) Measured transmission spectra of the critically coupled FRR through the bus (red) and drop (blue) ports as a function of the input laser detuning. 
    Two adjacent resonator modes are clearly visible and are separated by a free-spectral range of $\nu_\textrm{FSR}=(63.37\pm0.06)$ MHz.
    From top to bottom, the filtering bandwidth is increased through adjustment of both $\kappa_\textrm{drop}$ and $\kappa_\textrm{bus}$ while ensuring that the critical-coupling condition $\kappa_\textrm{bus}=\kappa_\textrm{0}+\kappa_\textrm{drop}$ is fulfilled. The latter ensures that the maximal on-resonance extinction (transmission) in the bus (drop) port is reached for a given filter bandwidth.}
\end{figure*}

\par Here, we present a simple and robust design for an ultra-narrow-bandwidth, fully tunable filter based on a fiber-ring resonator (FRR).
Its basic design is sketched in Fig. \ref{figCartoonSketch}.
The FRR is coupled to a waveguide that injects light into it and transmits the light field resulting from the interference between the input field and the cavity field.
When critically coupled, this realizes a two-port device that acts as a band-stop filter, see Fig. \ref{figCartoonSketch} (a), blocking resonant and transmitting non-resonant light.
When setting a bandwidth of $(1.80\pm0.02)$ MHz, our filter shows an extinction of $(20.5\pm0.7)$~dB in the NIR/VIS spectral range, thereby outperforming any commercially available band-stop filter to our knowledge.
Coupling a second waveguide to the FRR realizes an add-drop filter that allows one to recover the resonant light and direct it to an independent output.
In this add-drop configuration, the filter can also be operated as a band-pass, see Fig. \ref{figCartoonSketch} (b), and, when setting the bandwidth to $(3.00\pm0.02)$ MHz, it shows an extinction of ($20.1\pm0.1$)~dB and a transmission of $(38.4\pm0.2)\%$ at the band-pass output. 
In the context of quantum optics, these characteristics make our filter compatible with resolving sub-natural linewidth features of laser-cooled atoms and ions as well as dye molecules at cryogenic temperatures.
It is thus of great interest, e.g., for analyzing and tailoring the fluorescence light of these quantum emitters. \cite{ng22, Masters2023, Liu2024}

\par The experimental setup for the characterization of the FRR is shown in Fig. \ref{figExpSetup} (a).
The FRR is made by splicing together the terminations of two commercially available variable ratio directional couplers (F-CPL-830-N-FA, Newport). 
Each coupler consists of a pair of single-mode, non-polarization-maintaining optical fibers.
Light can be coupled into and out of the resonator through the bus and drop fiber, respectively.
The FRR is placed in a thermally and acoustically insulated box in order to limit the thermal drifts and mechanical vibrations.
The optical round trip length of the FRR can be varied by stretching the fiber with a piezoelectric actuator (PK4FA2H3P2, Thorlabs), which also allows us to actively stabilize the resonance frequency. 
The FRR exhibits two sets of polarization eigenmodes due to the birefringence of the optical fibers.
Using two in-line polarization controllers (CPC900, Thorlabs), the frequency splitting of the FRR eigenmodes can be set at will.
Depending on the desired operation, the FRR can be made polarization-sensitive (if the two sets of eigenmodes are spectrally separated) or polarization-independent (if the two sets of eigenmodes coincide).
For our characterization, we used a laser source that is stabilized at a wavelength of 852 nm and an output power of about 100 $\mu$W.
Two photodetectors record the output power at the bus and the drop port, respectively.
For the following analysis, we set the splitting of the polarization eigenmodes to approximately $\nu_\textrm{FSR}/4$ and match the polarization of the input light to one of the two sets of polarization eigenmodes.
By tuning the splitting ratio of the variable ratio directional couplers, we can adjust the loss rates of the FRR to the bus port ($\kappa_\textrm{bus}$) and the drop port ($\kappa_\textrm{drop}$).
The total loss rate of the FRR is thus given by $\kappa =\kappa_\textrm{0}+\kappa_\textrm{bus}+\kappa_\textrm{drop}$, where $\kappa_0$ is its intrinsic loss rate.
Fig. \ref{figExpSetup} (b) shows the transmission spectra at the bus and at the drop port for various settings of $\kappa$ under the condition of critical coupling.
An optimal point for filter operation is critical coupling, $\kappa_\textrm{bus}=\kappa_\textrm{0} + \kappa_\textrm{drop}$, at which, regardless of the finesse $\mathcal{F}$ of the resonator, the transmission to the bus port ideally is zero and the transmission to the drop port is maximal.
For $\mathcal{F} \gg 1$ the power transmission coefficients through the bus port and to the drop port are given by the expressions: \cite{Haus1984} 
\begin{equation}\label{eqTBus}
T_\textrm{bus} = \abs{\frac{\kappa_0+\kappa_\textrm{drop}-\kappa_\textrm{bus}+i \Delta} {\kappa_0+\kappa_\textrm{drop}+\kappa_\textrm{bus}+i \Delta}}^2,
\end{equation}
\begin{equation}\label{eqTDrop}
T_\textrm{drop} = \frac{4 \kappa_\textrm{drop} \kappa_\textrm{bus}}{\abs{\kappa_0+\kappa_\textrm{drop}+\kappa_\textrm{bus}+i \Delta}^2},
\end{equation}
where $\Delta$ is the detuning from the cavity resonance.
As is apparent from Eq. (1), $T_\textrm{bus}=0$ when $\kappa_\textrm{bus}=\kappa_\textrm{0} + \kappa_\textrm{drop}$.
\newline

\begin{figure}[t]
\centering
    \includegraphics[width=0.48\textwidth,keepaspectratio]{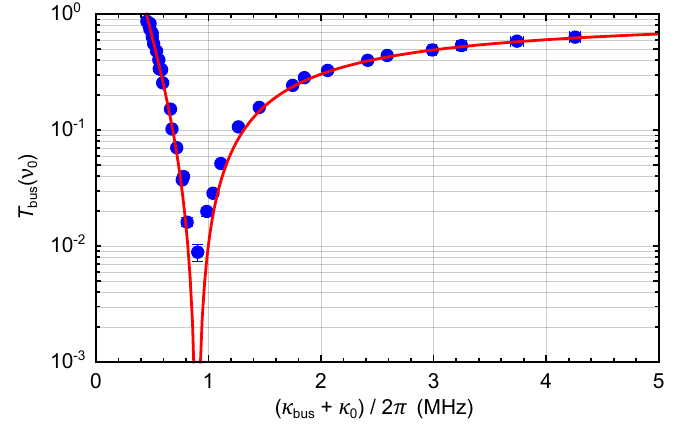}
    \caption{\label{figGraphTrKBusK0DropWithZoom} On-resonance power transmission through the bus port, $T_\textrm{bus}(\nu_0)$, as a function of the loss rate, $\kappa_\textrm{bus}+\kappa_\textrm{0}$, for $\kappa_\textrm{drop}=0$. 
    For each data point, $\kappa_\textrm{bus}+\kappa_\textrm{0}$ is obtained from a Lorentzian fit to the transmission spectrum around resonance, where the error bars represent the 1$\sigma$ uncertainty of the fit and are smaller than the size of the data points.
    The red solid line is a fit according to Eq. \ref{eqTBus}, with $\kappa_\textrm{0}$ as a free parameter.
    Its minimum indicates the point of critical coupling, where $\kappa_\textrm{bus}=\kappa_\textrm{0}$, from which an intrinsic loss rate of $\kappa_\textrm{0}/2\pi=(0.449\pm0.005)$ MHz is obtained.}
\end{figure}

\par We now study the on-resonance power transmission of the FRR through the bus port when it is operated as a band-stop filter.
Here, we keep $\kappa_\textrm{drop}$ fixed at 0 while varying $\kappa_\textrm{bus}$.
For each value of $\kappa_\textrm{bus}$, we fit an inverted Lorentzian to the transmission spectrum through the bus port.
From each fit, the linewidth and the resonance frequency of a cavity mode are extracted.
The total loss rate $\kappa$ is then obtained from the fitted linewidth, while the on-resonance power transmission $T_\textrm{bus}(\nu_0)$ is evaluated by averaging the transmission around the resonance frequency over a range that is approximately $5\%$ of the linewidth given by the fit, where the error bars are given by the standard deviation of the mean.
The resulting values are shown in Fig. \ref{figGraphTrKBusK0DropWithZoom} together with a fitting curve obtained from Eq. \ref{eqTBus} by setting $\kappa_\textrm{drop}=0$ and $\Delta=0$.
From the value of $\kappa$ at which the fitted curve reaches its minimum, corresponding to the critical coupling point, we estimate the intrinsic loss rate of the FRR to be $\kappa_\textrm{0}/2\pi=(0.449\pm0.005)$ MHz.
This corresponds to a total FWHM filter bandwidth of $(1.80\pm0.02)$ MHz and an unloaded finesse of $\mathcal{F}_0=70\pm1$. 
Considering that the length of the FRR ($3.257\pm0.003$ m) is much shorter than the inverse of the nominal optical attenuation coefficient for the used fiber (which exceeds 1 km), propagation losses are negligible.
Therefore, $\mathcal{F}_0$ is limited by the loss caused by the splices and by the variable ratio directional couplers. 
In addition, the experimental on-resonance transmission measured near critical coupling is $T_\textrm{bus}=(0.9\pm0.1)\%$, whereas the transmission in between two resonances is ${T_\textrm{bus}(\nu_0+\nu_\textrm{FSR}/2)}=(99.0\pm0.1)\%$. 
This corresponds to an extinction of $\abs{10\ log_{10} \frac{T_\textrm{bus}(\nu_0)}{T_\textrm{bus}(\nu_0+\nu_\textrm{FSR}/2)}}=(20.5\pm0.7)$ dB. 
This extinction is most likely limited by the linewidth of our laser source (approximately $100$ kHz, FWHM), rather than by the FRR itself.
\newline

\begin{figure}[hb]
\centering
    \includegraphics[width=0.48\textwidth,keepaspectratio]{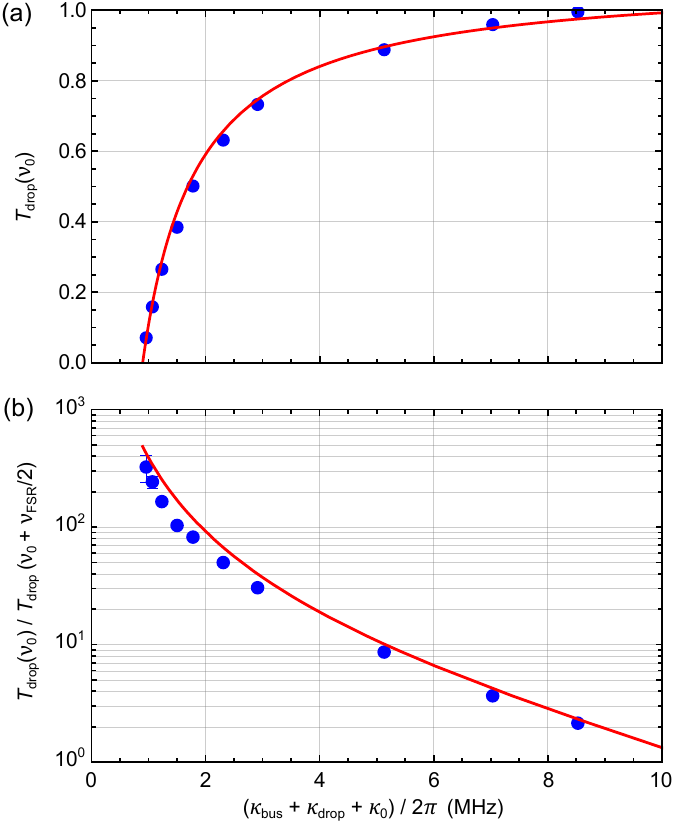}
\caption{\label{figTrDrop} (a) On-resonance transmission through the drop port, $T_\textrm{drop}(\nu_0)$, as a function of the total loss rate, $\kappa_\textrm{bus}+\kappa_\textrm{drop}+\kappa_\textrm{0}$, where the FRR is always set to critical coupling to the bus port ($\kappa_\textrm{bus}=\kappa_\textrm{0}+\kappa_\textrm{drop}$). (b) Extinction of the on-resonance transmission through the drop port as a function of the total loss rate. It is apparent that increasing the transmission comes at the cost of a reduced extinction because of the partial overlap between neighboring cavity modes. 
The total loss rate is evaluated by fitting the model represented by Eq. (\ref{eqTDropGen}) to the transmission spectra, where the error bars represent the 1$\sigma$ uncertainty of the fit and are smaller than the size of the data points. 
In (a) and (b), the transmission is evaluated by averaging the data around the resonance frequency and in between two resonances, respectively, and the error bars are given by the standard deviation of the mean.
The red lines are the predictions of the multi-mode cavity model.
}
\end{figure}

\par The second use case of the FRR is as a filter in the add-drop configuration.
In this case, the drop port causes additional losses, lowering the finesse $\mathcal{F}$ of the FRR.
For this reason, Eqs. (\ref{eqTBus}) and (\ref{eqTDrop}) need to be generalized in order to take into account the presence of adjacent longitudinal cavity modes (see Appendix A for further details). 
The generalized expressions were used to fit the transmission spectra at the drop port for various settings of the total loss rate $\kappa$, where $\kappa$ is the fit parameter. 
Fig. \ref{figTrDrop} (a) shows the on-resonance transmission $T_\textrm{drop}(\nu_0)$, whereas Fig. \ref{figTrDrop} (b) shows the extinction $T_\textrm{drop}(\nu_0)/T_\textrm{drop}(\nu_0+\nu_\textrm{FSR}/2)$ as a function of $\kappa$.
These values were evaluated from averaging the transmission around $\nu_0$ and around $\nu_0 + \nu_\textrm{FSR}$, respectively.
Our characterization shows that, for $\kappa/2\pi=(1.50\pm0.01)$ MHz, i.e., a FWHM filter bandwidth of $(3.00\pm0.02)$ MHz, the FRR has an extinction of $(20.1\pm0.1)$ dB or a peak-to-valley transmission ratio of about $100/1$. 
The transmission spectrum for this specific setting is represented in the first row of Fig. \ref{figExpSetup} (b).
These values are easily tunable and can be adapted to different requirements, realizing, e.g., a narrower bandwidth and a higher extinction, albeit at the cost of a lower on-resonance transmission through the drop port.

\par In conclusion, we have constructed and characterized an optical filter based on a fiber-ring resonator operable in an add-drop configuration, demonstrating its ultra-narrowband and high-extinction filtering capability.
When operated as a band-stop and as a band-pass filter, exemplary properties are summarized in Table \ref{tabS1}.

\begin{table}[h]
\caption{\label{tabS1} Summary of exemplary filter parameters in the two distinct operation modes.
    \textbf{BW}: bandwidth.
    \textbf{T($\nu_0$)}: on-resonance transmission.
    \textbf{{T($\nu_1$)}}: transmission in between two resonances.
    \textbf{EXT}: extinction.}
\begin{ruledtabular}
\begin{tabular}{ccccc}
\bf{} & \bf{BW} & \bf{T($\nu_0$)} &
\bf{T($\nu_1$)} & \bf{EXT}\\
\bf{} & (MHz) & (\%) & (\%) & (dB)\\
\hline
\bf{Band-stop} & $1.80(2)$ & $0.9(1)$ & $99.0(1)$ & $20.5(7)$\\
\bf{Band-pass} & $3.00(2)$ & $38.4(2)$ & $0.37(5)$ & $20.1(1)$\\	
\end{tabular}
\end{ruledtabular}
\end{table}

In principle, the filter bandwidth can be made narrower or broader by using a FRR of different length.
This would change the free-spectral range $\nu_\textrm{FSR}$, and since the finesse $\mathcal{F}_0$ is almost independent of the FRR length, its linewidth at critical coupling would change by the same amount.
Furthermore, the filter shows a very good passive stability, with a maximum passive frequency drift of approximately 1.2 MHz/minute (see Appendix B for further details).
The parameters of the demonstrated filter are widely tunable, where control over the filtering bandwidth, transmission, frequency and polarization are regulated by two separate fiber ports, a fiber stretcher and polarization controllers. 
Its ultra-narrow bandwidth makes it compatible with resolving spectroscopic features in the MHz-regime.
In particular, we recently used the filter to selectively suppress the coherent emission from a single optically trapped atom which was weakly driven with near-resonant light, thereby showing that the remaining incoherent emission exhibits bunched photon statistics. \cite{Masters2023}
Our filter design thus represents a viable alternative to current established resonator-based filters with potential applications beyond quantum optics, including laser stabilization, spectral noise suppression, polarimetry, and temperature or strain sensing. \cite{Mik2019, Fuderer2022, Bessin2019, Barrett2019}

\begin{acknowledgments}
We acknowledge funding by the Alexander von Humboldt Foundation in the framework of the Alexander von Humboldt Professorship endowed by the Federal Ministry of Education and Research, as well as by the European Commission under the project DAALI (No.899275). X. H. acknowledges a Humboldt Research Fellowship by the Alexander von Humboldt Foundation.\\
\end{acknowledgments}

%\section*{Author Declarations}
%The authors have no conflicts to disclose.

\section*{Data Availability Statement}

The data that support the findings of this study are available from the corresponding author upon reasonable request.

\appendix

\section{Resonator transmission for high losses}

When the finesse of the FRR is so low that the total loss rate $\kappa$ is no longer much smaller than the free-spectral-range $\nu_\textrm{FSR}$, consecutive cavity modes are not fully separated in the transmission spectrum. 
This occurs in particular when operating the FRR as a band-pass filter, where the addition of the drop fiber lowers the finesse and Eqs. (\ref{eqTBus}) and (\ref{eqTDrop}) need to be generalized to account for this effect.

\begin{figure}[h]
\centering
    \includegraphics[width=0.48\textwidth,keepaspectratio]{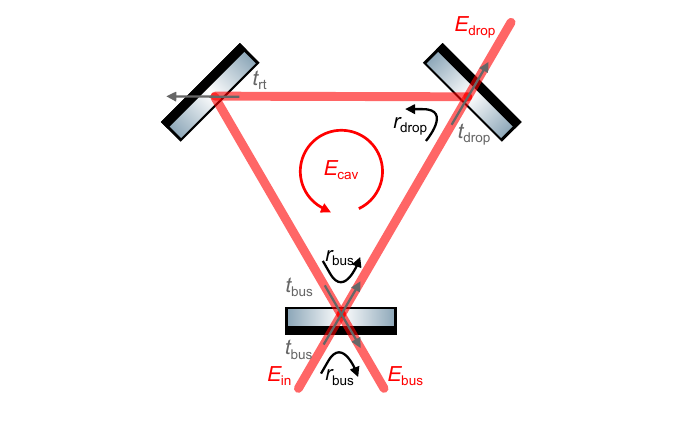}
    \caption{\label{figFRRSketch2} Sketch of the ring resonator model under consideration, featuring the notation used in the text.}
\end{figure}

The system can be modeled as a ring resonator consisting of three semi-reflective mirrors, one per each loss channel (bus, drop, and intrinsic losses), as shown in Fig. \ref{figFRRSketch2}.
We denote $r_\textrm{bus}$ and $t_\textrm{bus}$ as the amplitude reflection and transmission coefficients at the bus port, $r_\textrm{drop}$ and $t_\textrm{drop}$ as the amplitude reflection and transmission coefficients at the drop port, and $t_\textrm{rt}$ as the intrinsic amplitude transmission coefficient per round trip.
A self-consistent calculation of the electromagnetic fields yields:

\begin{equation}\label{eqTBusGen}
T_\textrm{bus} = \abs{\frac{E_\textrm{bus}}{E_\textrm{in}}}^2 = \abs{\frac{e^{\frac{i \Delta}{\nu_\textrm{FSR}}}r_\textrm{drop} t_\textrm{rt}-r_\textrm{bus}} {e^{\frac{i \Delta}{\nu_\textrm{FSR}}}r_\textrm{bus} r_\textrm{drop} t_\textrm{rt}-1}}^2,
\end{equation}
\begin{equation}\label{eqTDropGen}
T_\textrm{drop} = \abs{\frac{E_\textrm{drop}}{E_\textrm{in}}}^2 = \abs{\frac{t_\textrm{bus} t_\textrm{drop}} {-e^{\frac{i \Delta}{\nu_\textrm{FSR}}}r_\textrm{bus} r_\textrm{drop} t_\textrm{rt}+1}}^2,
\end{equation}

\noindent where $T_\textrm{bus}$ and $T_\textrm{drop}$ are the power transmission at the bus and the drop port, respectively. 
The introduced coefficients are related to the loss rates defined in the main text by $t_\textrm{rt}=\sqrt{1-\frac{2 \kappa_\textrm{0}}{\nu_\textrm{FSR}}}$, $r_\textrm{bus}=\sqrt{1-\frac{2 \kappa_\textrm{bus}}{\nu_\textrm{FSR}}}$, $r_\textrm{drop}=\sqrt{1-\frac{2 \kappa_\textrm{drop}}{\nu_\textrm{FSR}}}$.
In addition, the relation $r_\textrm{bus}^2+t_\textrm{bus}^2=r_\textrm{drop}^2+t_\textrm{drop}^2=1$ holds.
As a consequence, the critical coupling is reached when $r_\textrm{bus}=r_\textrm{drop} t_\textrm{rt}$ or, in terms of the loss rates:

\begin{equation}\label{eqCritCouplGen}
\kappa_\textrm{bus} = \frac{-2\kappa_\textrm{0}\kappa_\textrm{drop}+\kappa_\textrm{0}\nu_\textrm{FSR}+\kappa_\textrm{drop}\nu_\textrm{FSR}} {\nu_\textrm{FSR}}
\end{equation}

In order to infer the loss rate $\kappa$ corresponding to the data points shown in Fig. \ref{figTrDrop}, we fitted the respective transmission spectra with Eq. (\ref{eqTDropGen}) while fixing $\kappa_\textrm{0}$ to the value obtained from the analysis shown in Fig. \ref{figGraphTrKBusK0DropWithZoom}, fixing $\nu_\textrm{FSR}$ to the value obtained from an independent measurement, imposing the condition in Eq. (\ref{eqCritCouplGen}), and leaving solely $\kappa$ as a free parameter of the fit.
We note that Eqs. (\ref{eqTBusGen}) and (\ref{eqTDropGen}) reduce to Eqs. (\ref{eqTBus}) and (\ref{eqTDrop}) in the limiting case $\kappa\ll\nu_\textrm{FSR}$.

\section{Passive stability of the filter}

\begin{figure}[b]
\centering
    \includegraphics[width=0.48\textwidth,keepaspectratio]{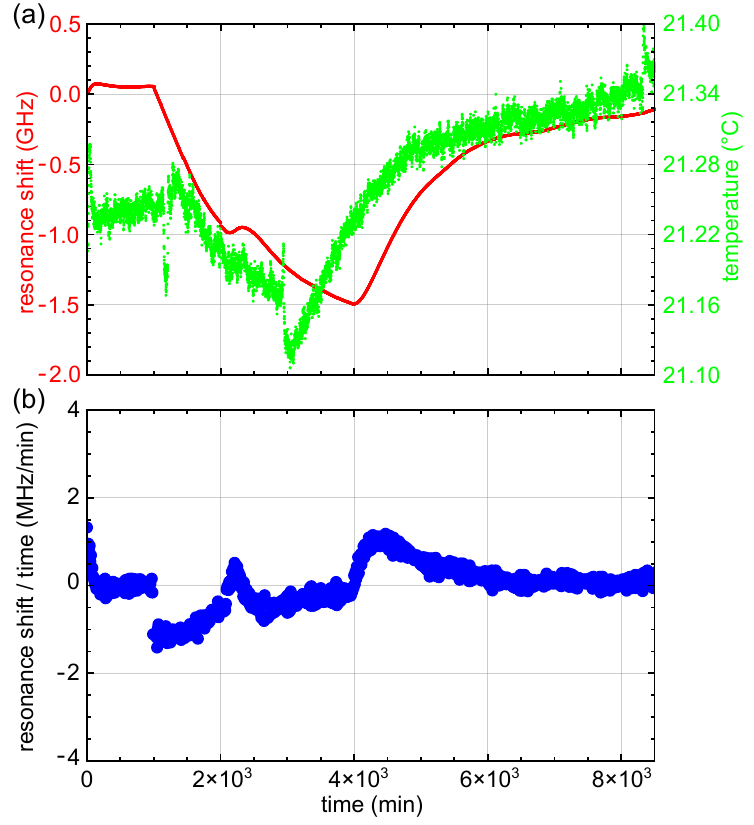}
\caption{\label{ResDriftComparModel} 
Long-term frequency drift of the FRR. 
(a) The red curve shows the resonance frequency of one of the FRR modes, measured once per minute. 
The green curve shows the simultaneously measured temperature in the insulating box.
(b) Inferred rate of frequency drift of the FRR.}
\end{figure}

\par In order to characterize the passive stability of the filter, we set it to critical coupling in the band-stop configuration without frequency stabilization and monitored the drift of the resonance frequency of one of its modes for about one week (around $10^4$ min). 
At the same time, we recorded the temperature inside the insulating box with a thermocouple.
The two measurement records are shown superposed in Fig. \ref{ResDriftComparModel} (a).
The observed non-monotonous drift of the resonance frequency shows correlations with the temperature, likely due to variations in the refractive index of the optical fiber. \cite{Fuderer2022}
We attribute the time shift between the temperature variations and the response of the FRR to the different thermal time constants of the optical board (where the FRR was placed) and the wooden wall of the insulating box (where the thermocouple was placed).
In Fig. \ref{ResDriftComparModel} (b), we show the time derivative of the resonance frequency drift.
Each data point is the average drift over 10 minutes, in order to account for the uncertainty on the evaluation of the resonance frequency (approximately 0.5 MHz). 
We observe that the maximum drift rate of the resonance frequency is around 1.2 MHz/minute, which can be counteracted by the active stabilization and thermal isolation of our setup.
As a reference, in \cite{Masters2023}, we used a sample and hold method to stabilize the resonance frequency of our FRR every 5 seconds. This ensures that the maximum drift is at most $2\%$ of the FWHM filter bandwidth. 

\section{Ring-down measurement of the intrinsic FRR loss rate}

\begin{figure}[h]
\centering
\includegraphics[width=0.48\textwidth,keepaspectratio]{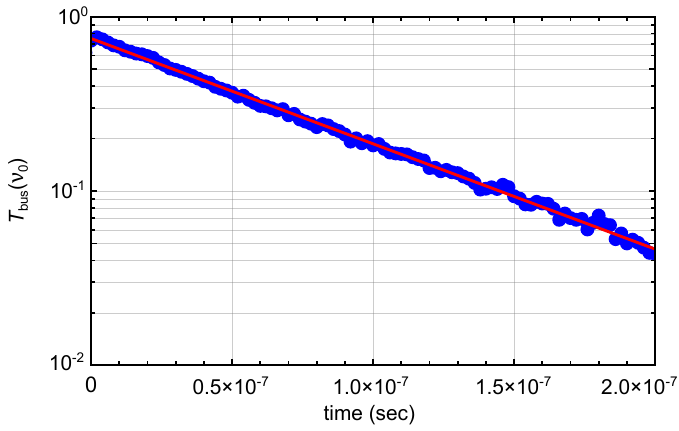}
\caption{\label{GraphExpDecay} 
Energy decay of the intra-cavity field when the filter is operated in the notch configuration at the critical coupling point.
At time $t=0$ the input field was switched off. Blue: experimental data. Red: exponential fit.}
\end{figure}

We also determine the intrinsic FRR loss rate $\kappa_\textrm{0}$ via a ring-down measurement.
To this end, we operate the filter in the notch configuration ($\kappa_\textrm{drop}=0$) close to the critical coupling point and launch resonant light through the bus fiber.
While monitoring the transmission at the bus port, we abruptly switch off the input field.
We observe that, following a steep increase, the transmission decreases exponentially with a characteristic time of $1/4\kappa_\textrm{0}$.
The fit in Fig. \ref{GraphExpDecay} yields an intrinsic loss rate of $\kappa_\textrm{0}/2\pi=(0.56\pm0.04)$ MHz, corresponding to a total filter bandwidth of $(2.23\pm0.04)$ MHz.
This value is slightly higher than that reported in the main text, likely due to a minor misadjustment in setting the critical coupling during the ring-down measurement. 

\bibliography{MainText}

\providecommand{\noopsort}[1]{}\providecommand{\singleletter}[1]{#1}%
\begin{thebibliography}{24}%
\makeatletter
\providecommand \@ifxundefined [1]{%
 \@ifx{#1\undefined}
}%
\providecommand \@ifnum [1]{%
 \ifnum #1\expandafter \@firstoftwo
 \else \expandafter \@secondoftwo
 \fi
}%
\providecommand \@ifx [1]{%
 \ifx #1\expandafter \@firstoftwo
 \else \expandafter \@secondoftwo
 \fi
}%
\providecommand \natexlab [1]{#1}%
\providecommand \enquote  [1]{``#1''}%
\providecommand \bibnamefont  [1]{#1}%
\providecommand \bibfnamefont [1]{#1}%
\providecommand \citenamefont [1]{#1}%
\providecommand \href@noop [0]{\@secondoftwo}%
\providecommand \href [0]{\begingroup \@sanitize@url \@href}%
\providecommand \@href[1]{\@@startlink{#1}\@@href}%
\providecommand \@@href[1]{\endgroup#1\@@endlink}%
\providecommand \@sanitize@url [0]{\catcode `\\12\catcode `\$12\catcode
  `\&12\catcode `\#12\catcode `\^12\catcode `\_12\catcode `\%12\relax}%
\providecommand \@@startlink[1]{}%
\providecommand \@@endlink[0]{}%
\providecommand \url  [0]{\begingroup\@sanitize@url \@url }%
\providecommand \@url [1]{\endgroup\@href {#1}{\urlprefix }}%
\providecommand \urlprefix  [0]{URL }%
\providecommand \Eprint [0]{\href }%
\providecommand \doibase [0]{http://dx.doi.org/}%
\providecommand \selectlanguage [0]{\@gobble}%
\providecommand \bibinfo  [0]{\@secondoftwo}%
\providecommand \bibfield  [0]{\@secondoftwo}%
\providecommand \translation [1]{[#1]}%
\providecommand \BibitemOpen [0]{}%
\providecommand \bibitemStop [0]{}%
\providecommand \bibitemNoStop [0]{.\EOS\space}%
\providecommand \EOS [0]{\spacefactor3000\relax}%
\providecommand \BibitemShut  [1]{\csname bibitem#1\endcsname}%
\let\auto@bib@innerbib\@empty
%</preamble>
\bibitem [{\citenamefont {Bass}(1995)}]{Bass1995}%
  \BibitemOpen
  \bibfield  {author} {\bibinfo {author} {\bibfnamefont {M.}~\bibnamefont
  {Bass}},\ }\href@noop {} {\emph {\bibinfo {title} {Handbook of Optics}}}\
  (\bibinfo  {publisher} {McGraw-Hill, Inc.},\ \bibinfo {year}
  {1995})\BibitemShut {NoStop}%
\bibitem [{\citenamefont {Macleod}(2010)}]{Macleod}%
  \BibitemOpen
  \bibfield  {author} {\bibinfo {author} {\bibfnamefont {H.~A.}\ \bibnamefont
  {Macleod}},\ }\href {\doibase 10.1201/9781420073034} {\emph {\bibinfo {title}
  {Thin-Film Optical Filters}}}\ (\bibinfo  {publisher} {CRC Press},\ \bibinfo
  {year} {2010})\BibitemShut {NoStop}%
\bibitem [{\citenamefont {Giacomo}, \citenamefont {Baumeister},\ and\
  \citenamefont {Jenkins}(1959)}]{giacomo59}%
  \BibitemOpen
  \bibfield  {author} {\bibinfo {author} {\bibfnamefont {P.}~\bibnamefont
  {Giacomo}}, \bibinfo {author} {\bibfnamefont {P.~W.}\ \bibnamefont
  {Baumeister}}, \ and\ \bibinfo {author} {\bibfnamefont {F.~A.}\ \bibnamefont
  {Jenkins}},\ }\bibfield  {title} {\enquote {\bibinfo {title} {On the limiting
  band width of interference filters},}\ }\href {\doibase
  10.1088/0370-1328/73/3/314} {\bibfield  {journal} {\bibinfo  {journal}
  {Proceedings of the Physical Society}\ }\textbf {\bibinfo {volume} {73}},\
  \bibinfo {pages} {480} (\bibinfo {year} {1959})}\BibitemShut {NoStop}%
\bibitem [{\citenamefont {Koonen}(2006)}]{Koonen2006}%
  \BibitemOpen
  \bibfield  {author} {\bibinfo {author} {\bibfnamefont {T.}~\bibnamefont
  {Koonen}},\ }\enquote {\bibinfo {title} {Fabry-p\'erot interferometer
  filters},}\ in\ \href {\doibase 10.1007/3-540-31770-8_7} {\emph {\bibinfo
  {booktitle} {Wavelength Filters in Fibre Optics}}},\ \bibinfo {editor}
  {edited by\ \bibinfo {editor} {\bibfnamefont {H.}~\bibnamefont {Venghaus}}}\
  (\bibinfo  {publisher} {Springer Berlin Heidelberg},\ \bibinfo {year}
  {2006})\ pp.\ \bibinfo {pages} {271--287}\BibitemShut {NoStop}%
\bibitem [{\citenamefont {Xue}\ \emph {et~al.}(2012)\citenamefont {Xue},
  \citenamefont {Tao}, \citenamefont {Sun}, \citenamefont {Hong}, \citenamefont
  {Zhuang}, \citenamefont {Luo}, \citenamefont {Chen},\ and\ \citenamefont
  {Guo}}]{Xue2012}%
  \BibitemOpen
  \bibfield  {author} {\bibinfo {author} {\bibfnamefont {X.}~\bibnamefont
  {Xue}}, \bibinfo {author} {\bibfnamefont {Z.}~\bibnamefont {Tao}}, \bibinfo
  {author} {\bibfnamefont {Q.}~\bibnamefont {Sun}}, \bibinfo {author}
  {\bibfnamefont {Y.}~\bibnamefont {Hong}}, \bibinfo {author} {\bibfnamefont
  {W.}~\bibnamefont {Zhuang}}, \bibinfo {author} {\bibfnamefont
  {B.}~\bibnamefont {Luo}}, \bibinfo {author} {\bibfnamefont {J.}~\bibnamefont
  {Chen}}, \ and\ \bibinfo {author} {\bibfnamefont {H.}~\bibnamefont {Guo}},\
  }\bibfield  {title} {\enquote {\bibinfo {title} {Faraday anomalous dispersion
  optical filter with a single transmission peak using a buffer-gas-filled
  rubidium cell},}\ }\href@noop {} {\bibfield  {journal} {\bibinfo  {journal}
  {Opt. Lett.}\ }\textbf {\bibinfo {volume} {37}},\ \bibinfo {pages}
  {2274--2276} (\bibinfo {year} {2012})}\BibitemShut {NoStop}%
\bibitem [{\citenamefont {Widmann}\ \emph {et~al.}(2018)\citenamefont
  {Widmann}, \citenamefont {Portalupi}, \citenamefont {Michler}, \citenamefont
  {Wrachtrup},\ and\ \citenamefont {Gerhardt}}]{Widmann2018}%
  \BibitemOpen
  \bibfield  {author} {\bibinfo {author} {\bibfnamefont {M.}~\bibnamefont
  {Widmann}}, \bibinfo {author} {\bibfnamefont {S.~L.}\ \bibnamefont
  {Portalupi}}, \bibinfo {author} {\bibfnamefont {P.}~\bibnamefont {Michler}},
  \bibinfo {author} {\bibfnamefont {J.}~\bibnamefont {Wrachtrup}}, \ and\
  \bibinfo {author} {\bibfnamefont {I.}~\bibnamefont {Gerhardt}},\ }\bibfield
  {title} {\enquote {\bibinfo {title} {Faraday filtering on the cs-d1-line for
  quantum hybrid systems},}\ }\href@noop {} {\bibfield  {journal} {\bibinfo
  {journal} {IEEE Phot. Techn. Lett.}\ }\textbf {\bibinfo {volume} {30}},\
  \bibinfo {pages} {2083--2086} (\bibinfo {year} {2018})}\BibitemShut {NoStop}%
\bibitem [{\citenamefont {Palittapongarnpim}, \citenamefont {MacRae},\ and\
  \citenamefont {Lvovsky}(2012)}]{palittapongarnpim12}%
  \BibitemOpen
  \bibfield  {author} {\bibinfo {author} {\bibfnamefont {P.}~\bibnamefont
  {Palittapongarnpim}}, \bibinfo {author} {\bibfnamefont {A.}~\bibnamefont
  {MacRae}}, \ and\ \bibinfo {author} {\bibfnamefont {A.~I.}\ \bibnamefont
  {Lvovsky}},\ }\bibfield  {title} {\enquote {\bibinfo {title} {Note: A
  monolithic filter cavity for experiments in quantum optics},}\ }\href@noop {}
  {\bibfield  {journal} {\bibinfo  {journal} {Rev. of Sci. Instrum.}\ }\textbf
  {\bibinfo {volume} {83}},\ \bibinfo {pages} {066101} (\bibinfo {year}
  {2012})}\BibitemShut {NoStop}%
\bibitem [{\citenamefont {Ahlrichs}\ \emph {et~al.}(2013)\citenamefont
  {Ahlrichs}, \citenamefont {Berkemeier}, \citenamefont {Sprenger},\ and\
  \citenamefont {Benson}}]{ahlrichs13}%
  \BibitemOpen
  \bibfield  {author} {\bibinfo {author} {\bibfnamefont {A.}~\bibnamefont
  {Ahlrichs}}, \bibinfo {author} {\bibfnamefont {C.}~\bibnamefont
  {Berkemeier}}, \bibinfo {author} {\bibfnamefont {B.}~\bibnamefont
  {Sprenger}}, \ and\ \bibinfo {author} {\bibfnamefont {O.}~\bibnamefont
  {Benson}},\ }\bibfield  {title} {\enquote {\bibinfo {title} {A monolithic
  polarization-independent frequency-filter system for filtering of photon
  pairs},}\ }\href@noop {} {\bibfield  {journal} {\bibinfo  {journal} {Appl.
  Phys. Lett.}\ }\textbf {\bibinfo {volume} {103}},\ \bibinfo {pages} {241110}
  (\bibinfo {year} {2013})}\BibitemShut {NoStop}%
\bibitem [{\citenamefont {Phillips}\ \emph {et~al.}(2020)\citenamefont
  {Phillips}, \citenamefont {Brash}, \citenamefont {McCutcheon}, \citenamefont
  {Iles-Smith}, \citenamefont {Clarke}, \citenamefont {Royall}, \citenamefont
  {Skolnick}, \citenamefont {Fox},\ and\ \citenamefont {Nazir}}]{phillips20}%
  \BibitemOpen
  \bibfield  {author} {\bibinfo {author} {\bibfnamefont {C.~L.}\ \bibnamefont
  {Phillips}}, \bibinfo {author} {\bibfnamefont {A.~J.}\ \bibnamefont {Brash}},
  \bibinfo {author} {\bibfnamefont {D.~P.}\ \bibnamefont {McCutcheon}},
  \bibinfo {author} {\bibfnamefont {J.}~\bibnamefont {Iles-Smith}}, \bibinfo
  {author} {\bibfnamefont {E.}~\bibnamefont {Clarke}}, \bibinfo {author}
  {\bibfnamefont {B.}~\bibnamefont {Royall}}, \bibinfo {author} {\bibfnamefont
  {M.~S.}\ \bibnamefont {Skolnick}}, \bibinfo {author} {\bibfnamefont {A.~M.}\
  \bibnamefont {Fox}}, \ and\ \bibinfo {author} {\bibfnamefont
  {A.}~\bibnamefont {Nazir}},\ }\bibfield  {title} {\enquote {\bibinfo {title}
  {Photon statistics of filtered resonance fluorescence (supplemental
  material)},}\ }\href@noop {} {\bibfield  {journal} {\bibinfo  {journal}
  {Phys. Rev. Lett.}\ }\textbf {\bibinfo {volume} {125}},\ \bibinfo {pages}
  {043603} (\bibinfo {year} {2020})}\BibitemShut {NoStop}%
\bibitem [{\citenamefont {Cordier}\ \emph {et~al.}(2023)\citenamefont
  {Cordier}, \citenamefont {Schemmer}, \citenamefont {Schneeweiss},
  \citenamefont {Volz},\ and\ \citenamefont {Rauschenbeutel}}]{cordier23}%
  \BibitemOpen
  \bibfield  {author} {\bibinfo {author} {\bibfnamefont {M.}~\bibnamefont
  {Cordier}}, \bibinfo {author} {\bibfnamefont {M.}~\bibnamefont {Schemmer}},
  \bibinfo {author} {\bibfnamefont {P.}~\bibnamefont {Schneeweiss}}, \bibinfo
  {author} {\bibfnamefont {J.}~\bibnamefont {Volz}}, \ and\ \bibinfo {author}
  {\bibfnamefont {A.}~\bibnamefont {Rauschenbeutel}},\ }\bibfield  {title}
  {\enquote {\bibinfo {title} {Tailoring photon statistics with an atom-based
  two-photon interferometer (supplemental materials)},}\ }\href@noop {}
  {\bibfield  {journal} {\bibinfo  {journal} {Phys. Rev. Lett.}\ }\textbf
  {\bibinfo {volume} {131}},\ \bibinfo {pages} {183601} (\bibinfo {year}
  {2023})}\BibitemShut {NoStop}%
\bibitem [{\citenamefont {Reed}(2008)}]{Reed}%
  \BibitemOpen
  \bibfield  {author} {\bibinfo {author} {\bibfnamefont {G.~T.}\ \bibnamefont
  {Reed}},\ }\href@noop {} {\emph {\bibinfo {title} {Silicon Photonics: The
  State of the Art}}}\ (\bibinfo  {publisher} {John Wiley \& Sons, Inc.},\
  \bibinfo {year} {2008})\BibitemShut {NoStop}%
\bibitem [{\citenamefont {Fox}, \citenamefont {Oates},\ and\ \citenamefont
  {Hollberg}(2003)}]{Fox2003}%
  \BibitemOpen
  \bibfield  {author} {\bibinfo {author} {\bibfnamefont {W.}~\bibnamefont
  {Fox}}, \bibinfo {author} {\bibfnamefont {C.~W.}\ \bibnamefont {Oates}}, \
  and\ \bibinfo {author} {\bibfnamefont {W.}~\bibnamefont {Hollberg}},\
  }\href@noop {} {\emph {\bibinfo {title} {1. Stabilizing diode lasers to
  high-finesse cavities}}},\ Vol.~\bibinfo {volume} {40}\ (\bibinfo
  {publisher} {Academic Press},\ \bibinfo {year} {2003})\ pp.\ \bibinfo {pages}
  {1--46}\BibitemShut {NoStop}%
\bibitem [{\citenamefont {Yoo}(1996)}]{Yoo1996}%
  \BibitemOpen
  \bibfield  {author} {\bibinfo {author} {\bibfnamefont {S.~J.~B.}\
  \bibnamefont {Yoo}},\ }\bibfield  {title} {\enquote {\bibinfo {title}
  {Wavelength conversion technologies for wdm network applications},}\
  }\href@noop {} {\bibfield  {journal} {\bibinfo  {journal} {J. Lightwave
  Technol.}\ }\textbf {\bibinfo {volume} {14}},\ \bibinfo {pages} {955--966}
  (\bibinfo {year} {1996})}\BibitemShut {NoStop}%
\bibitem [{\citenamefont {Hou}\ \emph {et~al.}(2023)\citenamefont {Hou},
  \citenamefont {Paokang}, \citenamefont {Manav}, \citenamefont {Ian},
  \citenamefont {Weichuan}, \citenamefont {Zhihong},\ and\ \citenamefont
  {Linran}}]{Hou2023}%
  \BibitemOpen
  \bibfield  {author} {\bibinfo {author} {\bibfnamefont {S.}~\bibnamefont
  {Hou}}, \bibinfo {author} {\bibfnamefont {C.}~\bibnamefont {Paokang}},
  \bibinfo {author} {\bibfnamefont {S.}~\bibnamefont {Manav}}, \bibinfo
  {author} {\bibfnamefont {B.}~\bibnamefont {Ian}}, \bibinfo {author}
  {\bibfnamefont {X.}~\bibnamefont {Weichuan}}, \bibinfo {author}
  {\bibfnamefont {L.}~\bibnamefont {Zhihong}}, \ and\ \bibinfo {author}
  {\bibfnamefont {F.}~\bibnamefont {Linran}},\ }\bibfield  {title} {\enquote
  {\bibinfo {title} {Programmable optical filter in thin-film lithium niobate
  with simultaneous tunability of extinction ratio and wavelength},}\
  }\href@noop {} {\bibfield  {journal} {\bibinfo  {journal} {ACS Photonics}\
  }\textbf {\bibinfo {volume} {10}},\ \bibinfo {pages} {3896--3900} (\bibinfo
  {year} {2023})}\BibitemShut {NoStop}%
\bibitem [{\citenamefont {Sun}\ \emph {et~al.}(2022)\citenamefont {Sun},
  \citenamefont {Chen}, \citenamefont {Yin}, \citenamefont {Ye}, \citenamefont
  {Luo}, \citenamefont {Ma}, \citenamefont {Jian}, \citenamefont {Shi},
  \citenamefont {Zhong}, \citenamefont {Zhang}, \citenamefont {Lin},\ and\
  \citenamefont {Li}}]{Sun2022}%
  \BibitemOpen
  \bibfield  {author} {\bibinfo {author} {\bibfnamefont {C.}~\bibnamefont
  {Sun}}, \bibinfo {author} {\bibfnamefont {Z.}~\bibnamefont {Chen}}, \bibinfo
  {author} {\bibfnamefont {Y.}~\bibnamefont {Yin}}, \bibinfo {author}
  {\bibfnamefont {Y.}~\bibnamefont {Ye}}, \bibinfo {author} {\bibfnamefont
  {Y.}~\bibnamefont {Luo}}, \bibinfo {author} {\bibfnamefont {H.}~\bibnamefont
  {Ma}}, \bibinfo {author} {\bibfnamefont {J.}~\bibnamefont {Jian}}, \bibinfo
  {author} {\bibfnamefont {Y.}~\bibnamefont {Shi}}, \bibinfo {author}
  {\bibfnamefont {C.}~\bibnamefont {Zhong}}, \bibinfo {author} {\bibfnamefont
  {D.}~\bibnamefont {Zhang}}, \bibinfo {author} {\bibfnamefont
  {H.}~\bibnamefont {Lin}}, \ and\ \bibinfo {author} {\bibfnamefont
  {L.}~\bibnamefont {Li}},\ }\bibfield  {title} {\enquote {\bibinfo {title}
  {Broadband and high-resolution integrated spectrometer based on a tunable
  fsr-free optical filter array},}\ }\href@noop {} {\bibfield  {journal}
  {\bibinfo  {journal} {ACS Photonics}\ }\textbf {\bibinfo {volume} {9}},\
  \bibinfo {pages} {2973--2980} (\bibinfo {year} {2022})}\BibitemShut {NoStop}%
\bibitem [{\citenamefont {Tong}\ \emph {et~al.}(2024)\citenamefont {Tong},
  \citenamefont {Xiong}, \citenamefont {Ding}, \citenamefont {Chen},
  \citenamefont {Qin}, \citenamefont {Chen}, \citenamefont {Tang},
  \citenamefont {Yu},\ and\ \citenamefont {Xu}}]{Tong2024}%
  \BibitemOpen
  \bibfield  {author} {\bibinfo {author} {\bibfnamefont {H.}~\bibnamefont
  {Tong}}, \bibinfo {author} {\bibfnamefont {Y.}~\bibnamefont {Xiong}},
  \bibinfo {author} {\bibfnamefont {H.}~\bibnamefont {Ding}}, \bibinfo {author}
  {\bibfnamefont {Z.}~\bibnamefont {Chen}}, \bibinfo {author} {\bibfnamefont
  {Z.}~\bibnamefont {Qin}}, \bibinfo {author} {\bibfnamefont {W.}~\bibnamefont
  {Chen}}, \bibinfo {author} {\bibfnamefont {J.}~\bibnamefont {Tang}}, \bibinfo
  {author} {\bibfnamefont {S.}~\bibnamefont {Yu}}, \ and\ \bibinfo {author}
  {\bibfnamefont {F.}~\bibnamefont {Xu}},\ }\bibfield  {title} {\enquote
  {\bibinfo {title} {Lithium niobate piezoelectric actuator-integrated fiber
  fabry-p\'erot tunable filter with ultrahigh speed and linearity},}\
  }\href@noop {} {\bibfield  {journal} {\bibinfo  {journal} {ACS Photonics}\
  }\textbf {\bibinfo {volume} {11}},\ \bibinfo {pages} {1574--1583} (\bibinfo
  {year} {2024})}\BibitemShut {NoStop}%
\bibitem [{\citenamefont {Ng}, \citenamefont {Chow},\ and\ \citenamefont
  {Kurtsiefer}(2022)}]{ng22}%
  \BibitemOpen
  \bibfield  {author} {\bibinfo {author} {\bibfnamefont {B.~L.}\ \bibnamefont
  {Ng}}, \bibinfo {author} {\bibfnamefont {C.~H.}\ \bibnamefont {Chow}}, \ and\
  \bibinfo {author} {\bibfnamefont {C.}~\bibnamefont {Kurtsiefer}},\ }\bibfield
   {title} {\enquote {\bibinfo {title} {Observation of the mollow triplet from
  an optically confined single atom},}\ }\href@noop {} {\bibfield  {journal}
  {\bibinfo  {journal} {Preprint at https://arxiv.org/abs/2208.06575v1}\ }
  (\bibinfo {year} {2022})}\BibitemShut {NoStop}%
\bibitem [{\citenamefont {Masters}\ \emph {et~al.}(2023)\citenamefont
  {Masters}, \citenamefont {Hu}, \citenamefont {Cordier}, \citenamefont
  {Maron}, \citenamefont {Pache}, \citenamefont {Rauschenbeutel}, \citenamefont
  {Schemmer},\ and\ \citenamefont {Volz}}]{Masters2023}%
  \BibitemOpen
  \bibfield  {author} {\bibinfo {author} {\bibfnamefont {L.}~\bibnamefont
  {Masters}}, \bibinfo {author} {\bibfnamefont {X.-X.}\ \bibnamefont {Hu}},
  \bibinfo {author} {\bibfnamefont {M.}~\bibnamefont {Cordier}}, \bibinfo
  {author} {\bibfnamefont {G.}~\bibnamefont {Maron}}, \bibinfo {author}
  {\bibfnamefont {L.}~\bibnamefont {Pache}}, \bibinfo {author} {\bibfnamefont
  {A.}~\bibnamefont {Rauschenbeutel}}, \bibinfo {author} {\bibfnamefont
  {M.}~\bibnamefont {Schemmer}}, \ and\ \bibinfo {author} {\bibfnamefont
  {J.}~\bibnamefont {Volz}},\ }\bibfield  {title} {\enquote {\bibinfo {title}
  {On the simultaneous scattering of two photons by a single two-level atom},}\
  }\href@noop {} {\bibfield  {journal} {\bibinfo  {journal} {Nat. Photon.}\
  }\textbf {\bibinfo {volume} {17}},\ \bibinfo {pages} {972--976} (\bibinfo
  {year} {2023})}\BibitemShut {NoStop}%
\bibitem [{\citenamefont {Liu}\ \emph {et~al.}(2024)\citenamefont {Liu},
  \citenamefont {Dall’Alba~Sandberg}, \citenamefont {Chan}, \citenamefont
  {Schrinski}, \citenamefont {Anyfantaki}, \citenamefont {Nielsen},
  \citenamefont {Larsen}, \citenamefont {Skalkin}, \citenamefont {Wang},
  \citenamefont {Midolo}, \citenamefont {Scholz}, \citenamefont {Wieck},
  \citenamefont {Ludwig}, \citenamefont {Sørensen}, \citenamefont {Tiranov},\
  and\ \citenamefont {Lodahl}}]{Liu2024}%
  \BibitemOpen
  \bibfield  {author} {\bibinfo {author} {\bibfnamefont {S.}~\bibnamefont
  {Liu}}, \bibinfo {author} {\bibfnamefont {O.~A.}\ \bibnamefont
  {Dall’Alba~Sandberg}}, \bibinfo {author} {\bibfnamefont {M.~L.}\
  \bibnamefont {Chan}}, \bibinfo {author} {\bibfnamefont {B.}~\bibnamefont
  {Schrinski}}, \bibinfo {author} {\bibfnamefont {Y.}~\bibnamefont
  {Anyfantaki}}, \bibinfo {author} {\bibfnamefont {R.~B.}\ \bibnamefont
  {Nielsen}}, \bibinfo {author} {\bibfnamefont {R.~G.}\ \bibnamefont {Larsen}},
  \bibinfo {author} {\bibfnamefont {A.}~\bibnamefont {Skalkin}}, \bibinfo
  {author} {\bibfnamefont {Y.}~\bibnamefont {Wang}}, \bibinfo {author}
  {\bibfnamefont {L.}~\bibnamefont {Midolo}}, \bibinfo {author} {\bibfnamefont
  {S.}~\bibnamefont {Scholz}}, \bibinfo {author} {\bibfnamefont {A.~D.}\
  \bibnamefont {Wieck}}, \bibinfo {author} {\bibfnamefont {A.}~\bibnamefont
  {Ludwig}}, \bibinfo {author} {\bibfnamefont {A.~S.}\ \bibnamefont
  {Sørensen}}, \bibinfo {author} {\bibfnamefont {A.}~\bibnamefont {Tiranov}},
  \ and\ \bibinfo {author} {\bibfnamefont {P.}~\bibnamefont {Lodahl}},\
  }\bibfield  {title} {\enquote {\bibinfo {title} {Violation of bell inequality
  by photon scattering on a two-level emitter},}\ }\href {\doibase
  10.1038/s41567-024-02543-8} {\bibfield  {journal} {\bibinfo  {journal} {Nat.
  Phys.}\ }\textbf {\bibinfo {volume} {20}},\ \bibinfo {pages} {1429--1433}
  (\bibinfo {year} {2024})}\BibitemShut {NoStop}%
\bibitem [{\citenamefont {Haus}(1984)}]{Haus1984}%
  \BibitemOpen
  \bibfield  {author} {\bibinfo {author} {\bibfnamefont {H.~A.}\ \bibnamefont
  {Haus}},\ }\href@noop {} {\emph {\bibinfo {title} {Waves and Fields in
  Optoelectronics}}}\ (\bibinfo  {publisher} {Prentice-Hall, Englewood Cliffs,
  New Jersey},\ \bibinfo {year} {1984})\BibitemShut {NoStop}%
\bibitem [{\citenamefont {Mik}\ \emph {et~al.}(2019)\citenamefont {Mik},
  \citenamefont {Sparkes}, \citenamefont {Perrella}, \citenamefont {Light},
  \citenamefont {Ng}, \citenamefont {Luiten},\ and\ \citenamefont
  {Ottaway}}]{Mik2019}%
  \BibitemOpen
  \bibfield  {author} {\bibinfo {author} {\bibfnamefont {J.~L.~H.}\
  \bibnamefont {Mik}}, \bibinfo {author} {\bibfnamefont {B.~M.}\ \bibnamefont
  {Sparkes}}, \bibinfo {author} {\bibfnamefont {C.}~\bibnamefont {Perrella}},
  \bibinfo {author} {\bibfnamefont {P.~S.}\ \bibnamefont {Light}}, \bibinfo
  {author} {\bibfnamefont {S.}~\bibnamefont {Ng}}, \bibinfo {author}
  {\bibfnamefont {A.~N.}\ \bibnamefont {Luiten}}, \ and\ \bibinfo {author}
  {\bibfnamefont {D.~J.}\ \bibnamefont {Ottaway}},\ }\bibfield  {title}
  {\enquote {\bibinfo {title} {High-transmission fiber ring resonator for
  spectral filtering of master oscillator power amplifiers},}\ }\href@noop {}
  {\bibfield  {journal} {\bibinfo  {journal} {OSA Continuum}\ ,\ \bibinfo
  {pages} {2487--2495}} (\bibinfo {year} {2019})}\BibitemShut {NoStop}%
\bibitem [{\citenamefont {Fuderer}\ \emph {et~al.}(2022)\citenamefont
  {Fuderer}, \citenamefont {Wang}, \citenamefont {Stuart}, \citenamefont
  {Hedges}, \citenamefont {Truscott},\ and\ \citenamefont
  {Hodgman}}]{Fuderer2022}%
  \BibitemOpen
  \bibfield  {author} {\bibinfo {author} {\bibfnamefont {L.~A.}\ \bibnamefont
  {Fuderer}}, \bibinfo {author} {\bibfnamefont {L.}~\bibnamefont {Wang}},
  \bibinfo {author} {\bibfnamefont {J.~S.}\ \bibnamefont {Stuart}}, \bibinfo
  {author} {\bibfnamefont {M.~P.}\ \bibnamefont {Hedges}}, \bibinfo {author}
  {\bibfnamefont {A.~G.}\ \bibnamefont {Truscott}}, \ and\ \bibinfo {author}
  {\bibfnamefont {S.~S.}\ \bibnamefont {Hodgman}},\ }\bibfield  {title}
  {\enquote {\bibinfo {title} {In-fibre temperature tuned fibre ring resonator
  for laser mode monitoring},}\ }\href@noop {} {\bibfield  {journal} {\bibinfo
  {journal} {Opt. Continuum}\ }\textbf {\bibinfo {volume} {1}},\ \bibinfo
  {pages} {306--314} (\bibinfo {year} {2022})}\BibitemShut {NoStop}%
\bibitem [{\citenamefont {Bessin}\ \emph {et~al.}(2019)\citenamefont {Bessin},
  \citenamefont {Perego}, \citenamefont {Staliunas}, \citenamefont {Turitsyn},
  \citenamefont {Kudlinski}, \citenamefont {Conforti},\ and\ \citenamefont
  {Mussot}}]{Bessin2019}%
  \BibitemOpen
  \bibfield  {author} {\bibinfo {author} {\bibfnamefont {F.}~\bibnamefont
  {Bessin}}, \bibinfo {author} {\bibfnamefont {A.}~\bibnamefont {Perego}},
  \bibinfo {author} {\bibfnamefont {K.}~\bibnamefont {Staliunas}}, \bibinfo
  {author} {\bibfnamefont {S.~K.}\ \bibnamefont {Turitsyn}}, \bibinfo {author}
  {\bibfnamefont {A.}~\bibnamefont {Kudlinski}}, \bibinfo {author}
  {\bibfnamefont {M.}~\bibnamefont {Conforti}}, \ and\ \bibinfo {author}
  {\bibfnamefont {A.}~\bibnamefont {Mussot}},\ }\bibfield  {title} {\enquote
  {\bibinfo {title} {Gain-through-filtering enables tuneable frequency comb
  generation in passive optical resonators},}\ }\href@noop {} {\bibfield
  {journal} {\bibinfo  {journal} {Nat. Commun.}\ }\textbf {\bibinfo {volume}
  {10}},\ \bibinfo {pages} {4489} (\bibinfo {year} {2019})}\BibitemShut
  {NoStop}%
\bibitem [{\citenamefont {Barrett}\ \emph {et~al.}(2019)\citenamefont
  {Barrett}, \citenamefont {Barter}, \citenamefont {Stuart}, \citenamefont
  {Yuen},\ and\ \citenamefont {Kuhn}}]{Barrett2019}%
  \BibitemOpen
  \bibfield  {author} {\bibinfo {author} {\bibfnamefont {T.~D.}\ \bibnamefont
  {Barrett}}, \bibinfo {author} {\bibfnamefont {O.}~\bibnamefont {Barter}},
  \bibinfo {author} {\bibfnamefont {D.}~\bibnamefont {Stuart}}, \bibinfo
  {author} {\bibfnamefont {B.}~\bibnamefont {Yuen}}, \ and\ \bibinfo {author}
  {\bibfnamefont {A.}~\bibnamefont {Kuhn}},\ }\bibfield  {title} {\enquote
  {\bibinfo {title} {Polarization oscillations in birefringent emitter-cavity
  systems},}\ }\href {\doibase 10.1103/PhysRevLett.122.083602} {\bibfield
  {journal} {\bibinfo  {journal} {Phys. Rev. Lett.}\ }\textbf {\bibinfo
  {volume} {122}},\ \bibinfo {pages} {083602} (\bibinfo {year}
  {2019})}\BibitemShut {NoStop}%
\end{thebibliography}%

\end{document}